\documentclass[apj]{emulateapj}

\begin{document}

\title{The Dynamics of Stellar Coronae Harboring Hot-jupiters II. A Space Weather Event on A Hot-jupiter}

\author{O. Cohen\altaffilmark{1}, V.L. Kashyap\altaffilmark{1}, J.J. Drake\altaffilmark{1},  
I.V. Sokolov\altaffilmark{2}, T.I. Gombosi\altaffilmark{2}}

\altaffiltext{1}{Harvard-Smithsonian Center for Astrophysics, 60 Garden St. Cambridge, MA 02138}
\altaffiltext{2}{Center for Space Environment Modeling, University of Michigan, 2455 Hayward St., 
Ann Arbor, MI 48109}

\begin{abstract}
We carry out a numerical simulation depicting the effects of a Coronal Mass Ejection (CME) event on a close-in giant planet in an extrasolar system. 
We drive the CME in a similar manner as in simulations of space weather events on Earth. The simulation 
includes the planetary orbital motion, which leads to the forming of a comet-like planetary magnetotail which is oriented almost perpendicular 
to the direction of propagation of the CME. The combination of this feature and the fact that the CME does not expand much by 
the time it reaches the planet leads to a unique CME-magnetosphere interaction, where the CME itself is highly affected 
by the presence of the planetary magnetosphere. This change in the CME properties throughout the event cannot be estimated 
by simple, analytic calculations. We find that the planet is well-shielded from CME penetration, even for 
a relatively weak intrinsic magnetic field. The planetary angular momentum loss associated with such an event is negligible 
compared to the total planetary angular momentum. We also find that the energy which is deposited in the magnetosphere is much 
higher than in the case of the Earth, and our simulation suggests there is a large-scale change in the orientation of the 
magnetosphere-ionosphere current system during the CME event. 
\end{abstract}

\keywords{stars: coronae - planet-star interactions - Sun: coronal mass ejections (CMEs)}


\section{INTRODUCTION}
\label{sec:Intro}

Coronal Mass Ejections (CMEs) are sudden releases of ionized gas from the stellar surface into interplanetary space. 
These explosions are triggered by a mechanism that is not yet fully understood, but it is thought to involve 
slow storage of magnetic energy that is quickly released by means of magnetic reconnection.  In the solar case, following 
the eruption, CMEs are accelerated in the corona, some to high terminal speeds of up to 3000~km~s$^{-1}$, where their 
magnetic integrity (commonly referred to as the  ``magnetic cloud'') is maintained as they are propagating to large 
interplanetary distances \citep[for a general review on the initiation, evolution, and 
propagation of CMEs, see][]{Forbes06}. By the time it reaches 1~AU, the interplanetary CME
would have expanded and reached a size much bigger than the magnetosphere of the Earth 
\citep{Forsyth06}. As a result, the ambient solar wind conditions in the vicinity of the Earth are replaced 
by the plasma conditions carried by the CME.  In some cases, when the orientation of the CME magnetic field 
is opposite to that of the Earth, a geomagnetic storm is driven by the energy exchange 
between the magnetic fields. This involves particle acceleration and an 
increase in radiation, as well as other phenomena such as increases in ring currents, and the appearance of 
the aurorae at high latitudes \citep{RussellKivelson95,Gombosi99}.  These geomagnetic phenomena, their 
impact and hazards, as well as the effort to predict them, have been 
collectively known by the umbrella phrase ``Space Weather'' \citep{Schwenn06,Pulkkinen07}. 

A large number of exoplanetary systems have been observed since the mid-1990's \citep{mayor95, exoplanet95, exoplanets03}. 
Until recently, most observed exoplanets were of the ``hot-jupiter'' type.  This class of 
planets are Jupiter-size gas giants, found in close-in orbits at distances as small as 0.01~AU from their host star 
\citep{exoplanet95, exoplanets03}.   Planets with sizes of Saturn, Uranus, Neptune, Super-earth, and even Earth have been 
recently observed by new instruments that are particularly designed for planet searches, such as {\it Kepler} 
\citep{KEPLER10}, {\it CoRoT} \citep{Corot09}, and {\it MOST} \citep{MOST03}.  Due to the proximity of hot-jupiters 
to their host star, it is possible that they can be located within the Alfv\'en point, at which the stellar wind becomes super-Alfv\'enic. 
Therefore, if a hot-jupiter has a substantial internal magnetic field, it can affect the stellar corona and the 
star via Star-Planet (magnetic) Interaction (SPI). The consequences of SPI have been discussed 
based on observations \citep{shkolnik03, shkolnik05a,shkolnik05b,shkolnik08,kashyap08,saar08,Pillitteri10,Poppenhaeger10}, 
theoretical models \citep{cuntz00,Ip04,lanza08,lanza09,lanza10}, and numerical simulations 
\citep{preusse06,lipatov05,Johansson09,cohen09b,Cohen10c}. The dominant scenario for SPI is a disruption of the 
ambient coronal magnetic structure by the planetary magnetic field due to the planetary orbital motion and via 
magnetic reconnection, as well as particle acceleration along field lines that connect the planet and the star 
towards the stellar surface. 

The proximity to the host star also causes close-in planets to suffer from high doses 
of X-ray and EUV radiation, which can affect their evolution, erosion, and atmospheric escape rate 
\citep[e.g.][]{Lammer03,Penz08,Yelle08,murray-clay09}.  In addition, transient CMEs might contribute to 
planetary erosion, especially if the planetary magnetic field is not strong enough to oppose the dynamic pressure of 
the CME.  It is commonly assumed that since stellar flare rates increase with stellar magnetic activity \citep[e.g.][]{Telleschi05} 
and CMEs are associated with flares \citep{gudel07}, the CME rate should be high 
for active stars.  Planetary erosion from such host stellar activity could be particularly enhanced. 

This issue is particularly important for planets orbiting M-dwarfs.  Since these stars have low luminosities, their habitable zone, 
which  covers the planetary distances where the surface temperature 
of a planet allows liquid water to exist \citep{Kasting93}, is located very close to the star.
Current planet detection methods are well-suited to find such planets which can potentially sustain life \citep{Scalo07}. 
The chance of habitability might be reduced, however, 
due to the high level of activity of M-dwarfs. \cite{Khodachenko07} and \cite{Lammer07} have studied the effect of CME erosion,
CME induced ion pickup, and extreme EUV radiation on the atmospheres of terrestrial planets orbiting an M class star.  They 
have used averaged scaled values of stellar CMEs based on solar CME statistics, as well as other scaling laws to characterize the M-type  
star and the stellar CME rate. They also determined the magnetospheric standoff distance and quantified the planetary magnetic moment 
necessary to shield the planet from CME erosion. 

The work presented in \cite{Khodachenko07} and \cite{Lammer07} is based on scaling 
an {\it Earth-like} interaction between planets and CMEs to a close-in orbit scenario. 
Here we present a full, three-dimensional, time-dependent, 
Magnetohydrodynamic (MHD) simulation of the interaction of a CME with the magnetosphere of a close-in hot-jupiter. 
The ultimate goal of our simulation is to study how different this interaction is from the case of the Earth due 
to the different magnetospheric structure, the interaction taking place partly in a sub-Alfv\'enic 
region, and the CME not being fully developed by the time it reaches the planet.  We also use 
the simulation to constrain the CME penetration depth for a particular planetary magnetic field strength and investigate the loss of 
planetary angular momentum as a result of the interaction.

We describe our numerical approach, the specifications of the planetary magnetosphere, and the CME initiation in \S\ref{sec:Model}. 
We present the simulation results in \S\ref{sec:Results} and discuss their consequences on close-in planets in 
\S\ref{sec:Discussion}. Finally, we summarize our conclusions in \S\ref{sec:Conclusions}.


\section{NUMERICAL SIMULATION}
\label{sec:Model}

\subsection{Numerical Approach}
\label{sec:Approach}

In the numerical simulation presented here, we study the response of the magnetic field of a close-in planet to a CME event 
using a single-fluid MHD model. The simulation includes both the stellar wind, the CME, and the planet. The planet 
is described by boundary values for the density and temperature, and by its magnetic field. In this work, 
we do not include a full description of the planetary atmosphere and magnetosphere, their response to the extreme 
stellar radiation, or detailed magnetospheric/ionospheric electrodynamics. Here we present a first step towards 
a more physical description of magnetosphere-CME interaction of close-in planets, which could be achieved 
in principle, by coupling the model for the stellar corona with a model for the planetary magnetosphere using 
a numerical framework \citep{toth05}.

Our work here comprises three ingredients: 1) the ambient stellar wind; 
2) the planet; and 3) the CME. We use the BATS-R-US global MHD model and its adaptation for the Solar Corona 
\citep{powell99,toth05,cohen07} to perform a simulation that includes all three components. This model has been 
adapted to simulations performed for stellar coronae, as well as exoplanetary systems \citep{Cohen09a,cohen09b,Cohen10c}.  
In addition, this model has been often used to simulate Sun-to-Earth 
space weather events in the solar system \citep[e.g.][]{manchester04,manchester08,lugaz07,cohen08a}, as well 
as studying the evolution of CMEs in the solar corona \citep[e.g.][]{Liu08,Jacobs09,cohen09c,Cohen10d}. 

The model solves the single-fluid set of conservation laws 
for mass, momentum, magnetic induction, and energy (ideal MHD equations):
\begin{eqnarray}
&\frac{\partial \rho}{\partial t}+\nabla\cdot(\rho \mathbf{u})=0,&  \nonumber \\
&\frac{\partial (\rho\mathbf{u})}{\partial t}+
\nabla\cdot\left(
\rho\mathbf{u}\mathbf{u}+pI+\frac{B^2}{2\mu_0}I-\frac{\mathbf{B}\mathbf{B}}{\mu_0}
\right) = \rho\mathbf{g},& \nonumber \\
&\frac{\partial \mathbf{B}}{\partial t}+
\nabla\cdot(\mathbf{u}\mathbf{B}-\mathbf{B}\mathbf{u})=0, &  \\
&\frac{\partial }{\partial t}\left(
\frac{1}{2}\rho u^2+\frac{1}{\Gamma-1}p+\frac{B^2}{2\mu_0}
 \right)+ &
 \nonumber \\
&
\nabla\cdot\left(
\frac{1}{2}\rho u^2\mathbf{u}+\frac{\Gamma}{\Gamma-1}p\mathbf{u}+
\frac{(\mathbf{B}\cdot\mathbf{B})\mathbf{u}-\mathbf{B}(\mathbf{B}\cdot\mathbf{u})}{\mu_0}
\right)+E_\gamma=\rho(\mathbf{g}\cdot\mathbf{u}), \nonumber &
\label{MHD}
\end{eqnarray}
in the inertial frame or in the rotating frame (where the centrifugal force is added). Here 
$\rho$, $\mathbf{u}$, $\mathbf{B}$, and $p$ are the plasma mass density, velocity, magnetic field, and 
thermal pressure, respectively. $\mu_0$ is the vacuum permeability, $g$ is the gravitational acceleration, 
and $\Gamma=1.5$ is the ratio of specific heats (polytropic index) at large distances from the Sun. $E_\gamma$ 
is an energy source term that depends on the local value of polytropic index. This term is responsible for the 
acceleration of the ambient stellar wind, and it is described in detail in Section~\ref{sec:Wind} below. 

\subsection{Ambient Stellar Wind}
\label{sec:Wind}

The ambient stellar wind is obtained in a semi-empirical manner based on an observed relation between 
the terminal solar wind speed, $u_{sw}$, and the magnetic flux tube expansion factor, $f_s$, from which the wind is originated 
\citep{cohen07}. Once an initial, potential (non-MHD) magnetic field distribution is specified from a surface magnetic field map, 
the value of the expansion factor for each magnetic 
field line is calculated, hence $u_{sw}$ from each field line can be determined, as well as the total kinetic energy per unit mass of the 
plasma, $u^2_{sw}/2$. Using Bernoulli integral, this energy can be used to specify the stellar surface value of the 
polytropic index, $\gamma_0$, assuming the boundary values for the density and pressure are known:
\begin{equation}
\frac{1}{2}u^2_{sw}-\frac{GM_\odot}{R_\odot}=\frac{\gamma_0}{\gamma_0-1}\frac{p_0}{\rho_0},
\end{equation}
with $G$, $M_\odot$, and $R_\odot$ being the gravitational constant, the stellar mass, and stellar radius, respectively. 
We then can define the energy source term:
\begin{equation}
E_\gamma=\frac{p}{\gamma-1}-\frac{p}{\Gamma-1},
\end{equation}
where $1<\gamma\le \Gamma$ is a radial function of $\gamma_0$. It can be seen that for smaller values of $\gamma$ (smaller expansion of 
the magnetic flux tube), the energy source is larger, and so is the acceleration of the wind. At large distances from the Sun 
($\sim 10R_\odot$), $\gamma\rightarrow\Gamma$, and $E_\gamma \rightarrow 0$.

The main advantage of this approach over the use of an ad-hoc wind solutions imposed on the magnetic field 
(the Parker wind solution for example; \citealt{parker58}) is that here the topology of the wind depends on the 
observed magnetic field structure \citep[e.g.,][]{Phillips95}. This model takes into account the dependency of the wind structure on the 
large-scale magnetic topology in the solar/stellar corona and produces a bi-modal wind with regions of slow, denser wind, 
(which originates from the open/closed field boundary where the expansion factor is large), and regions of fast, 
less dense wind (which originates from open field regions where the expansion factor is small) \citep{McComas07}. 

In order to avoid complexity and isolate the planet-CME 
interaction, we set the stellar magnetic field to be a dipole with an equatorial field strength of 2.5G aligned with the 
rotation axis of the star.  Other stellar parameters used here are matched to the observed stellar parameters of 
HD~189733, with stellar radius, $R_\star=0.76R_\odot$, stellar mass, $M_\star=0.2M_\odot$, and stellar rotation period, 
$\Omega_\star=11.95d$ \citep{exoplanet95,exoplanets03}. The ambient stellar wind 
is solar-like with terminal speeds ranging between $265\;km\;s^{-1}$ and $800\;km\;s^{-1}$ \citep[see][]{cohen09b,Cohen10c}.

\subsection{Modeling the Planet}
\label{sec:planet} 

\begin{figure*}[h!]
\centering
\includegraphics[width=2.5in]{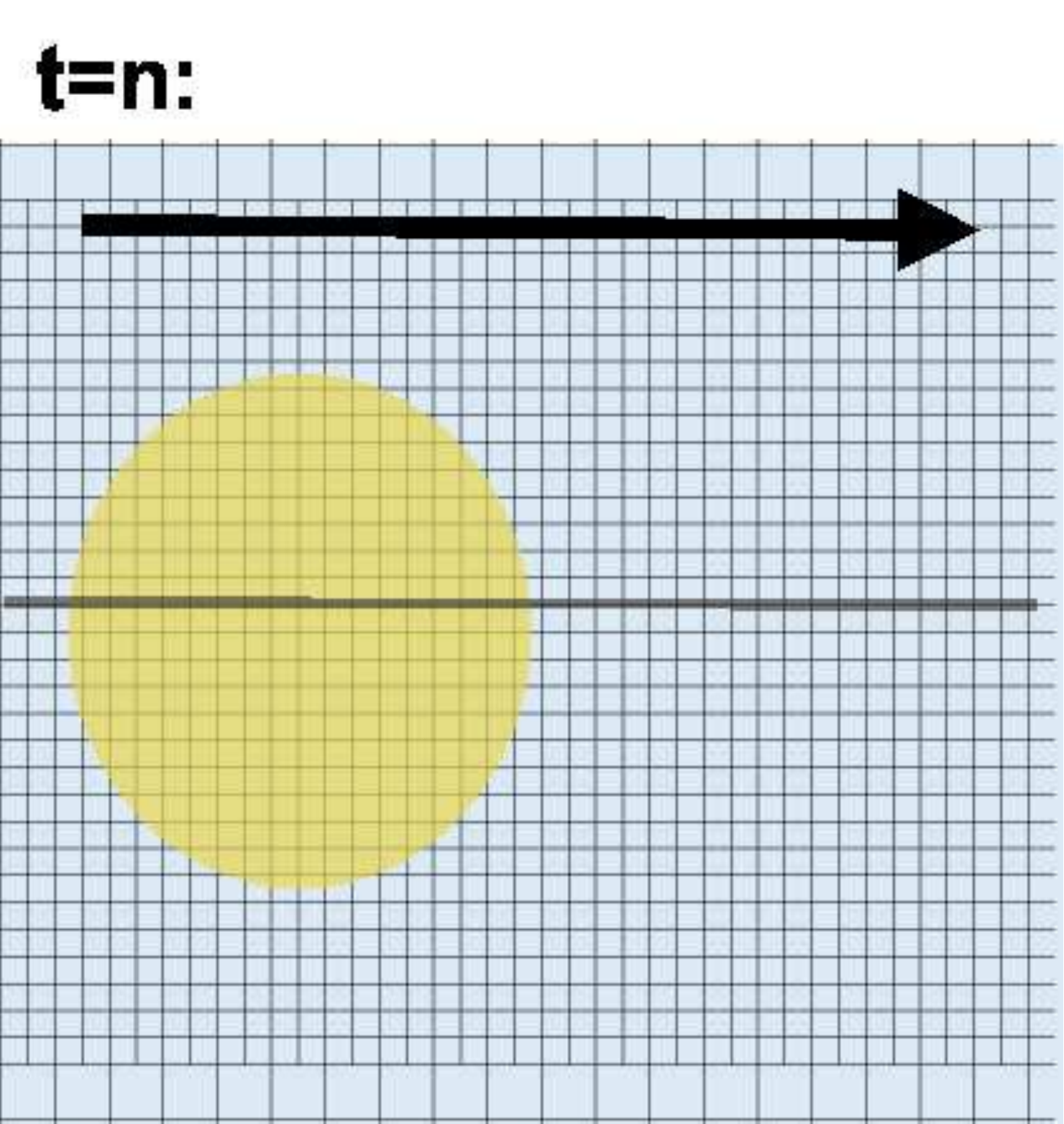}
\includegraphics[width=2.5in]{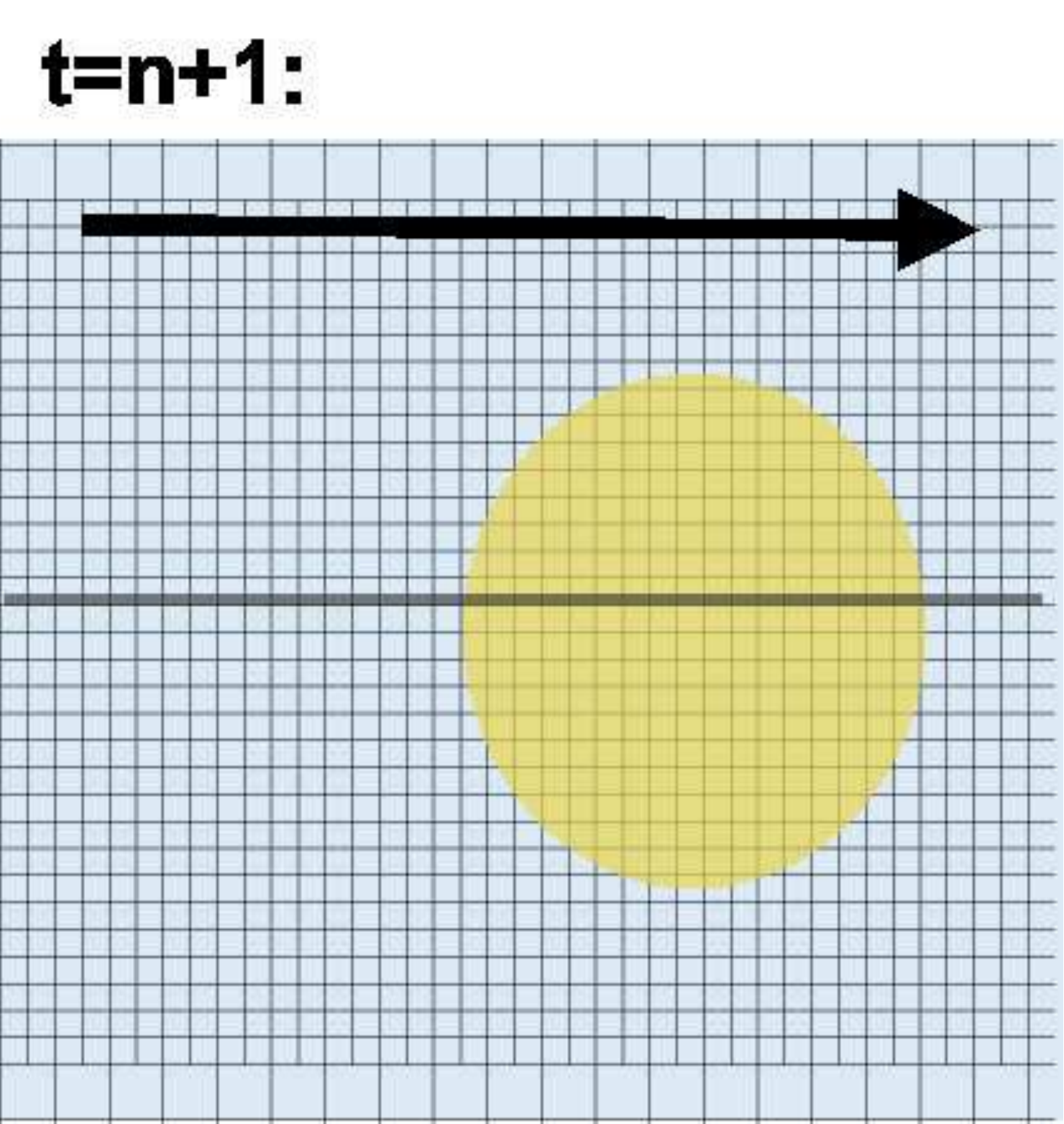}\\
\includegraphics[width=5.in]{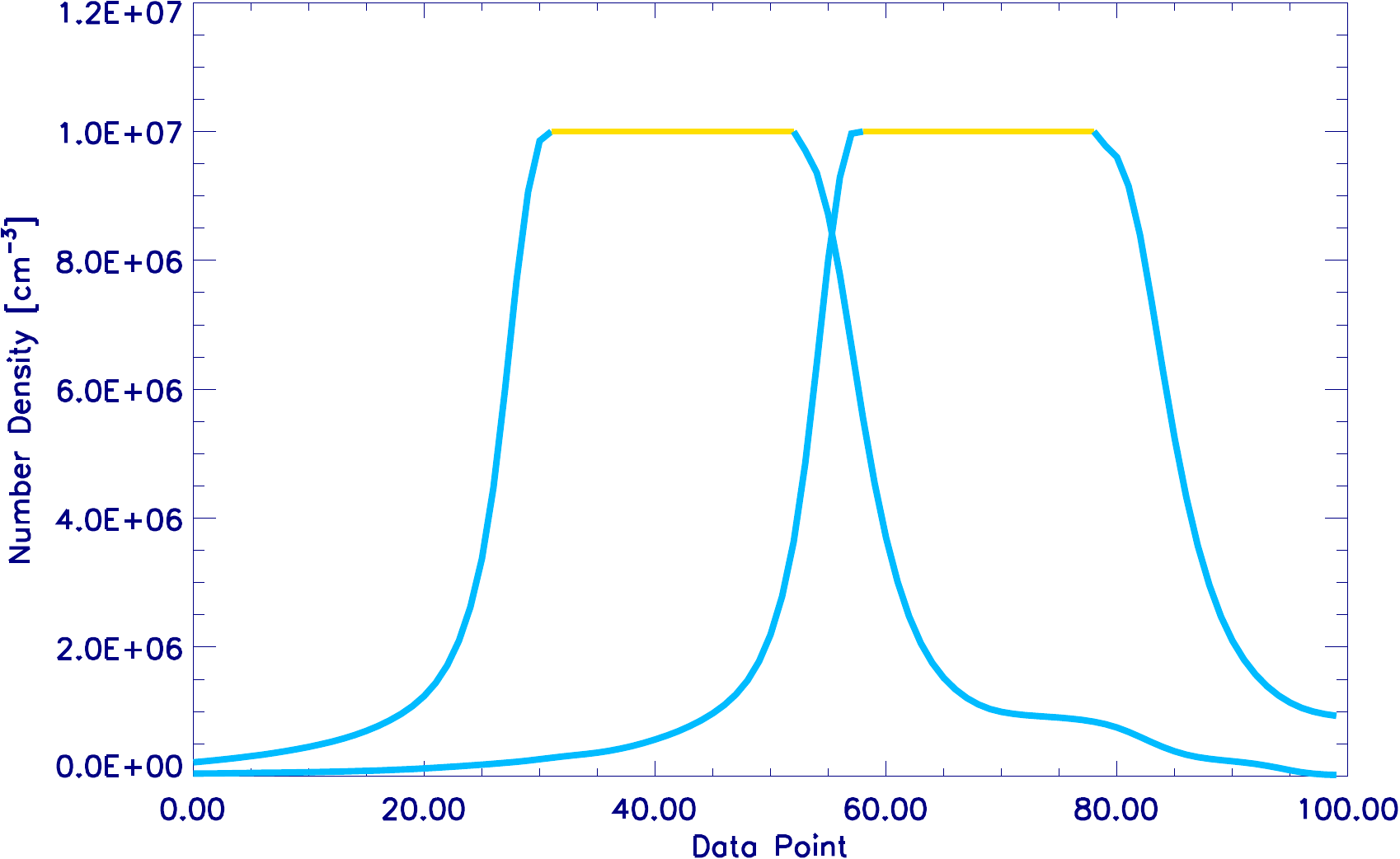}
\caption{Top - the change in cells definition as the planet's position is modified illustrated with the actual grid used in the simulation. 
At t=n, cells that are inside the body (colored in yellow) hold the fix boundary values, while cells outside the body (colored in blue) are 
updated via the MHD solution. 
At t=n+1, the body moves to its new position, so cells that were inside the body are now outside and are updated via the MHD solution, while 
some cells move inside the body and become ``body cells''. Bottom - a line extraction of the number density across the body at t=n and t=n+1 (along the gray line 
shown on the left). It can be seen that the high boundary value is moving with the planetary position.}
\label{fig:f1}
\end{figure*}

In order to model the planet, we impose an additional boundary condition in the simulation domain which 
is constrained by the planetary surface density, temperature, and magnetic field in a similar manner as in \cite{cohen09b} 
and \cite{Cohen10c}. At each time-step, the coordinates of this boundary (or planetary body) are updated based on the 
planetary orbit. Grid cells inside the body are defined as ``ghost cells'' or ``boundary cells'' and are forced to 
have the boundary values. The solution is updated given the particular set of ``boundary cells'' at a particular 
time-step. At each time step, the planet's location is updated, and boundary cells that no longer overlap the planet are 
returned to being a ``regular cell'' and their values are updated according to the MHD solution. This process is illustrated 
in Figure~\ref{fig:f1}. 
Here, the second boundary condition for the planet 
is set at a semi-major axis of $a=8.8R_\star$ and with a radius of 
$R_p=0.2R_\star\approx 1.5R_J$, where $R_J$ is Jupiter's radius. The boundary conditions for the base density 
and temperature are $n_p=10^7~cm^{-3}$ and $T_p=10^4~k$, respectively. 

Here we do not include the planetary gravity since several tests have shown that the effect of the small planetary mass on the solution is negligible. This is due to the 
fact that the planet is small compared to the simulation domain, so that the planetary scale height cannot be captured 
as also other  physical features, important on a planetary scale. The region near the planet is magnetically dominated, and so stress balance
between the magnetosphere and wind/CME does not depend on density (through thermal pressure, which requires a more realistic model for the planet). Therefore, the density gradient near the planet is dominated by the 
difference between the boundary density value and the ambient density of the stellar corona, as well as the gradient in planetary 
magnetic pressure. This density gradient is more moderate than that in reality, and it shows, for example, a drop of less than order of magnitude between 
$R_p$ and $2R_p$, while in reality one should expect a drop 2-3 orders of magnitude based on the planetary parameters. 

Planetary rotation is omitted; even though it could be implemented via the boundary condition for the velocity 
inside the second body (currently, the boundary condition for the planetary velocity is of the orbital velocity). We choose to ignore rotation in this 
simulation in order not to apply more complexity to a numerical boundary, which already has a delicate stability due to its tiny size. 
We assume that close-in planets do not rotate as fast as Jupiter due to their spindown by tidal locking processes \citep{sanches-lavega04}, 
and that the planetary magnetospheric dynamics is Earth-like (dominated by the solar wind) and not Jupiter-like 
(dominated by planetary rotation) \citep{RussellKivelson95}. 

\begin{figure*}[h!]
\centering
\includegraphics[width=6.in]{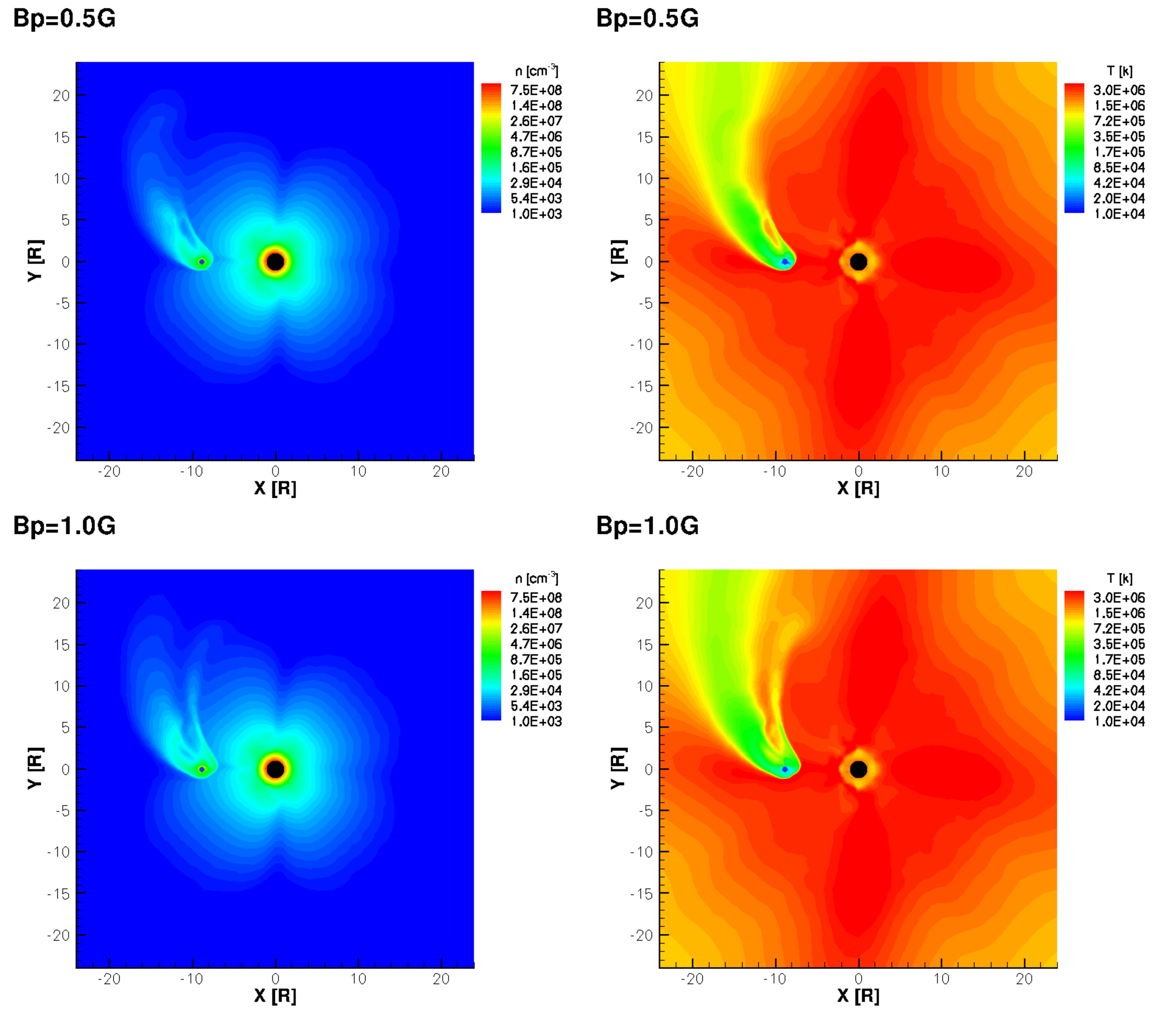}
\caption{The number density (left) and temperature (right) at the initial pre-eruption stage are shown for a slice in the equatorial plane in the simulation. 
The weak planetary field case (Case A) is shown in the top panels and the strong field case (Case B) is shown in the bottom panels. The star and the planet are 
shown as solid spheres.}
\label{fig:f2}
\end{figure*}

Since our goal is to quantify the planetary magnetic field strength necessary to shield the planet from CME events, 
we study two cases. One with a weak equatorial dipole field strength of $0.5~G$ (Case A hereafter) and one with a stronger equatorial 
dipole field strength 
of $1~G$ (Case B hereafter). Jupiter's equatorial field is about $4.3~G$. We start the simulation by letting the wind solution 
relax to a steady-state with a stationary (tidally locked) 
planet and then turn on the planetary orbital motion. The orbital motion is obtained by 
updating the coordinates of the planetary boundary at each time-step. This technique and its implications for the 
dynamics of stellar coronae harboring close-in planets is described in \cite{Cohen11}. We allow the 
simulation to evolve for half an orbit so that it does not include any perturbations generated by the 
initiation of the orbital motion. We then use this solution as our initial condition and launch the CME 
as described below (\S\ref{sec:CME}).  Figure~\ref{fig:f2} shows the distribution of the number density 
and temperature at this pre-eruption state over the equatorial plain of the simulation domain. The 
relatively high temperature of $\sim 10^5\;K$ in the planetary tail (comparing to the boundary value of $10^4\;K$) is due to the 
fact that it contains hot coronal plasma that is being trapped inside the planetary magnetosphere as it is sweeping through the corona.

\subsection{CME Initiation}
\label{sec:CME}

We initiate the CME by superimposing an unstable, semi-circular flux rope based on the analytical model by \cite{titov99} 
on top of the ambient ``initial'' solution described above (\S\ref{sec:planet}) \citep{roussev03}. In simulations of real 
solar CME events, the flux rope properties are matched to fit the observed properties of the source active region and its inversion 
line. The free energy provided to the CME is controlled by an additional toroidal field in order to produce the observed linear speed 
of the CME. Here, for the sake of definiteness, we use the parameters (except for orientation) matched to a CME event that occurred on May 13, 2005. This was a 
typical solar Halo CME event which gained a linear speed of about $1700~km~s^{-1}$ \citep{gopal09}, and is a Solar, Heliospheric, 
and INterplanetary Environment (SHINE\footnote{SHINE is the NSF branch for solar and heliospheric physics})  campaign event.  SHINE campaign 
events are picked in order to promote a community effort to study a particular CME event from different aspects (observations, theory, 
and numerical modeling). Here we investigate the impact of this CME on the exoplanet, and in particular, we 
study the CME penetration depth for a particular planetary field strength. Therefore, we choose to set the flux-rope 
orientation to be exactly opposite to the planetary dipole orientation. This way, the CME penetration is maximized, supported 
by magnetic reconnection between the CME and the planetary magnetic fields. 

\subsection{General Simulation Setup}
\label{sec:General}
We run the simulation for 6 hours after the CME onset time. By this time, the CME reaches the edges of the simulation domain. 
Here, we focus on the early part of the simulation when the CME reaches the planet, approximately 30 minutes after initiation. 
We omit later stages when the the planetary magnetosphere relaxes back to its initial state after the CME has passed, and in fact, 
the CME is never turned off.  
The simulation here is done with a model for the stellar corona.  Therefore, at this point we are unable to study
the planetary magnetospheric response to the CME in detail. In order to maintain the CME integrity and preventing numerical diffusion of the 
CME magnetic field, we specify the non-uniform grid to have very 
high-resolution around the source active region on the stellar surface with $\Delta x=10^{-3}R_\star$, and in the space between 
the star and the planet with $\Delta x=5\cdot10^{-2}R_\star$. The grid size around the planet and along its orbit is 
$\Delta x=1.1\cdot10^{-2}R_\star=0.055R_p$.


\section{RESULTS}
\label{sec:Results}

\subsection{CME-Planet Interaction}
\label{sec:Interaction}

\begin{figure*}[h!]
\centering
\includegraphics[width=6.in]{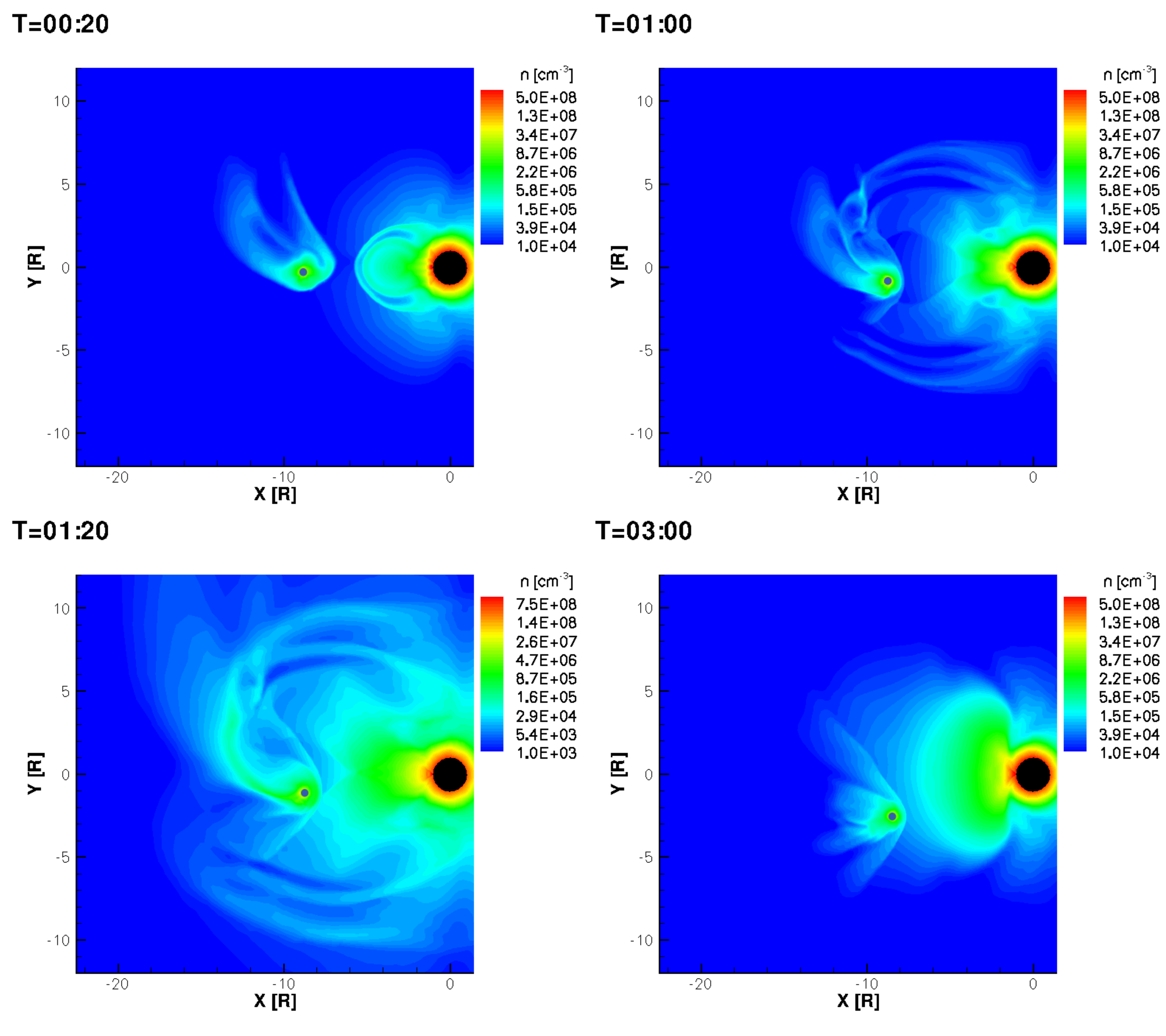}
\caption{Number density values in the equatorial plane, displayed at various times of the simulation, $t=$ 00:20h, 01:00h, 01:20h, and 03:00h.}
\label{fig:f3}
\end{figure*}

\begin{figure*}[h!]
\centering
\includegraphics[width=6.in]{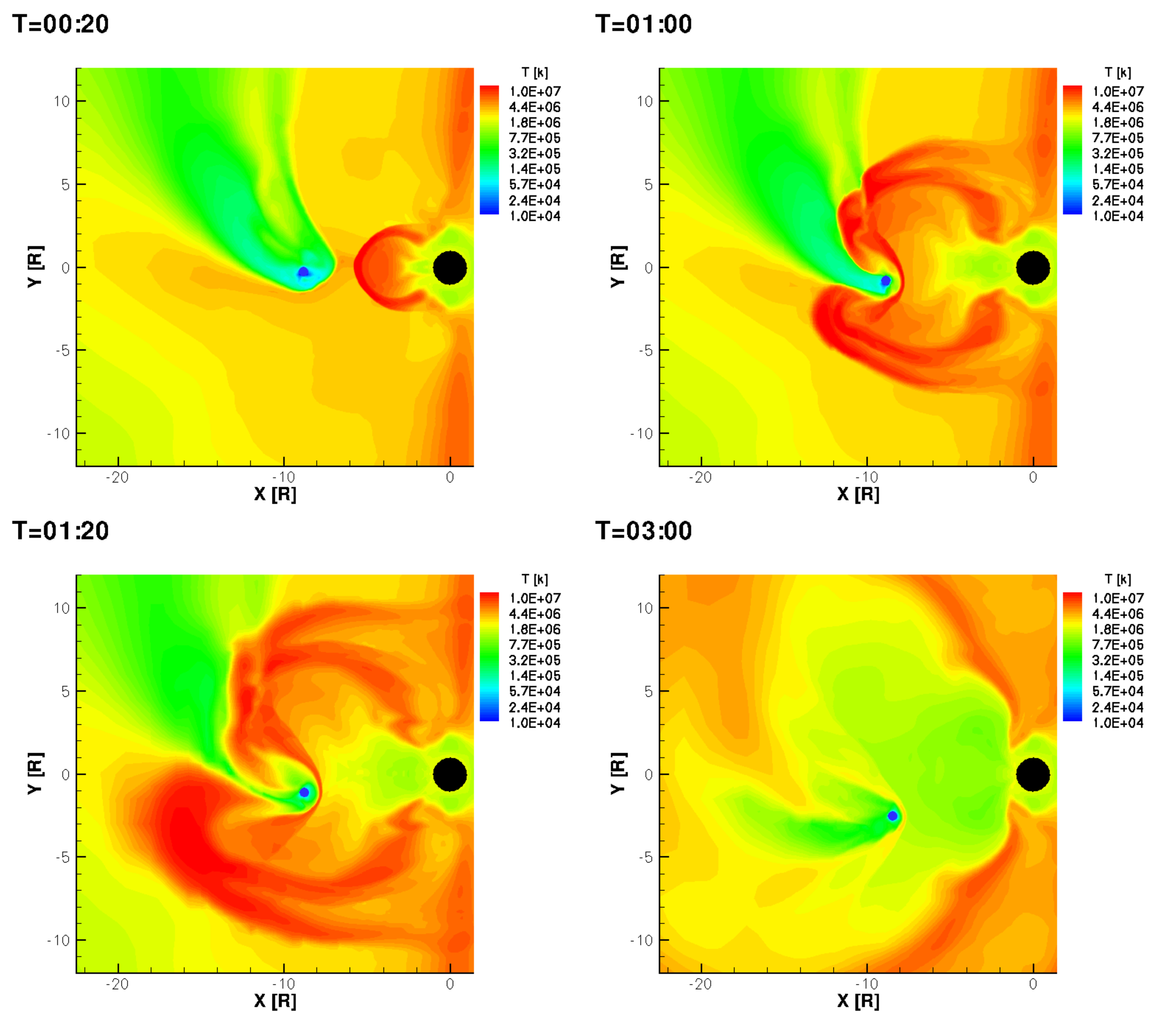}
\caption{Same as Figure~\ref{fig:f3} but for temperature.}
\label{fig:f4}
\end{figure*}

\begin{figure*}[h!]
\centering
\includegraphics[width=6.in]{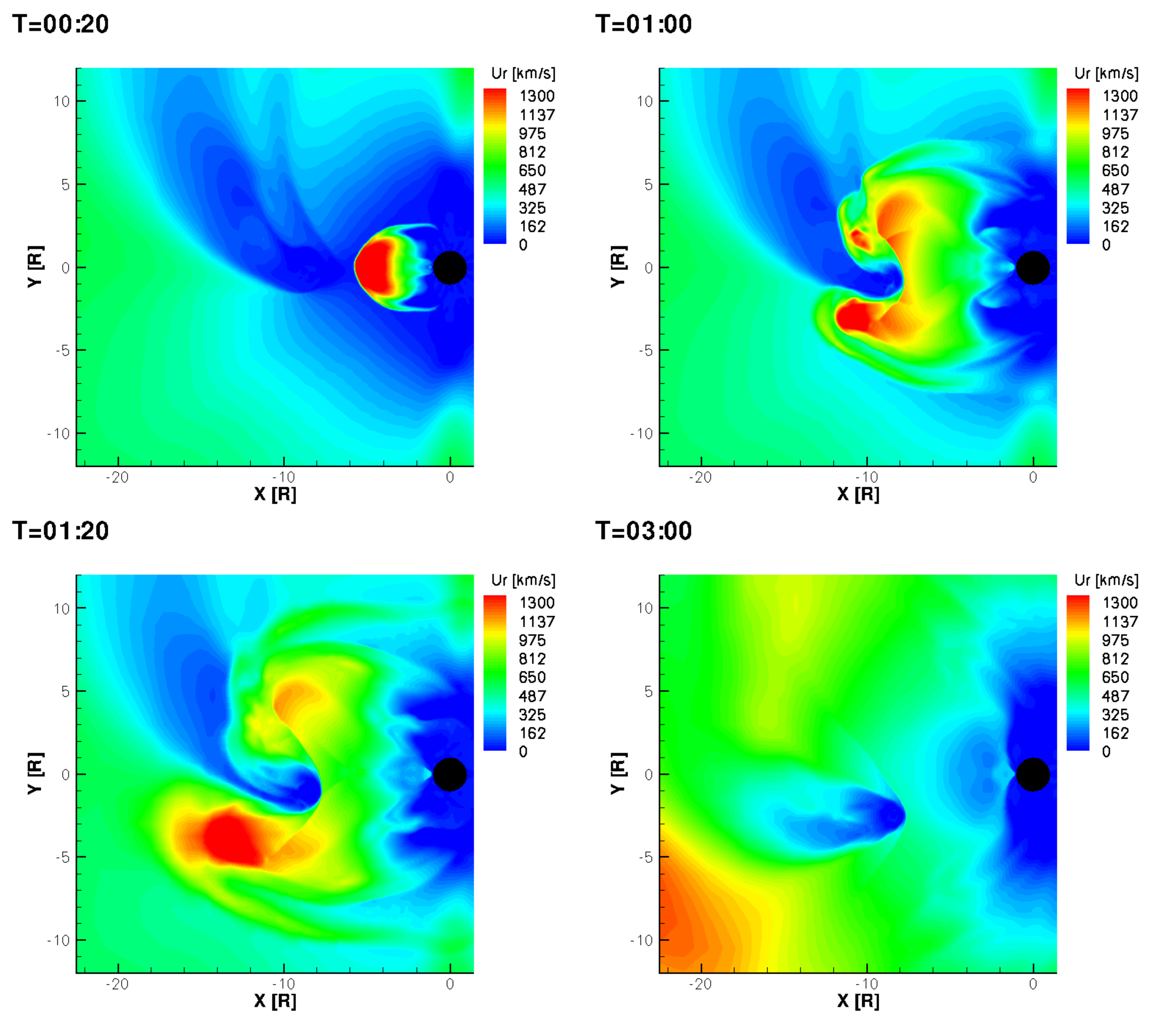}
\caption{Same as Figure~\ref{fig:f3} but for radial speed away from the star.}
\label{fig:f5}
\end{figure*}

Figures~\ref{fig:f3}-\ref{fig:f5} show the number density, temperature, and radial speed respectively, on the equatorial plain,
at t = 00:20, 01:00, 01:20, and 03:00 hours for Case B. An animation of these frames, as well as an animation of the three-dimensional 
magnetic field evolution can be found in the online version of the paper 
({\sl CaseB\_n.mov, CaseB\_T.mov, CaseB\_U.mov, and ESP\_CME\_3D.mov}), 
The large-scale coronal structure looks similar for Case A; 
the differences between the two cases are described later as we touch upon the more detailed aspects of the results. 

At t = 00:20h, the CME can be seen as it approaches 
the planetary magnetosphere. The CME front is characterized by a dense, hot plasma. When a CME is propagating through the solar corona, 
it expands and accelerates while pushing the ambient coronal plasma outward and interacting with its magnetic field.  In our simulation, a similar 
scenario is obtained until the CME reaches 
the nose of the planetary magnetosphere. Then, the CME is deflected and adiabatically expands around the magnetosphere front and envelopes it from the 
sides (noticeable at t = 01:00h). The CME front is slowed down and the temperature at its front drops.  Later on, at t = 01:20h, 
the CME reaches the night side of the planetary magnetosphere, while detaching from part of the planetary magnetic tail.  At this point, 
the size of the planetary magnetosphere, characterized by the low temperature bubble near the planet, is significantly decreased.  Finally, at t = 03:00, 
the magnetosphere starts to recover from 
the initial impact by the CME, it gets larger, and its tail is stretched in the anti-stellar direction. This is due to the ambient plasma, 
which is now dominated by the radial propagation and expansion of the CME. 

\subsection{CME Penetration Depth}
\label{sec:CMEPenetration}

The cold bubble around the planet is persistent throughout the simulations, as seen in
Figures~\ref{fig:f3}-\ref{fig:f5} for Case~B with a stronger planetary magnetic field, indicating that the CME never reaches the planetary surface directly.  
In order to quantify the penetration depth of the CME towards the planetary surface more precisely, we extract three spheres around the 
planet at height of $h = 0.5$, 1, and $2R_p$.  We calculate the total mass flux, $F_p=\int \rho \mathbf{u_p}\cdot\mathbf{da}$, 
flowing through each surface throughout the simulation, where $\mathbf{u_p}$ is the outward radial velocity in the planetary frame of reference, and 
$\mathbf{da}$ is a surface element of the sphere. Since the density at the surface of the planet is higher than the density of the ambient coronal 
plasma, the total mass flux over the spheres at the chosen distances should be positive (a net mass loss from the planet). Therefore, 
a change in the sign of $F$ means that the CME has overcome the positive planetary outflow and penetrates through the particular sphere, changing the 
net flux through the surface from positive to negative. 

\begin{figure*}[h!]
\centering
\includegraphics[width=6.in]{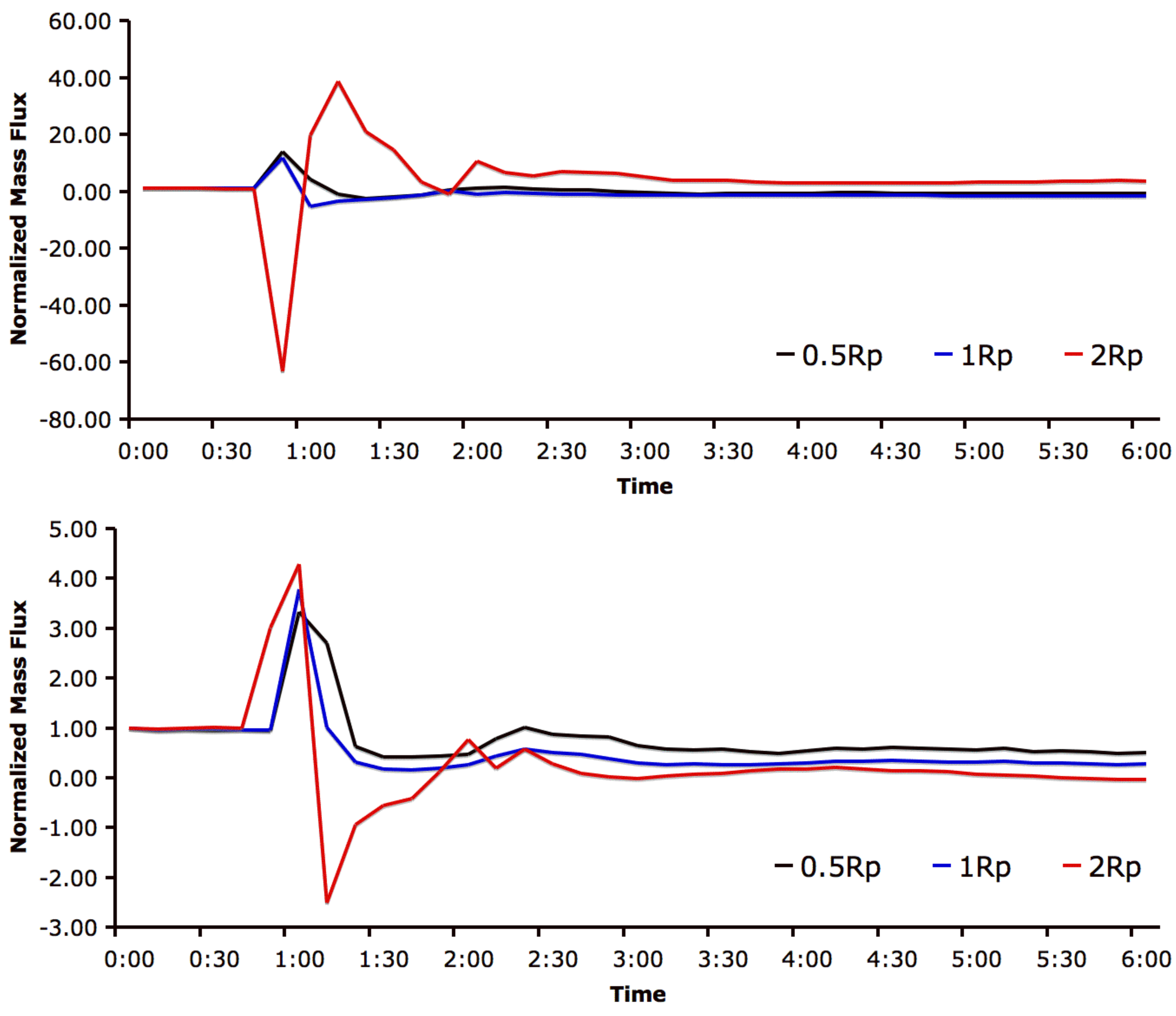}
\caption{Total mass flux through spherical surfaces with radii $r=0.5R_p$ (black curve), $r=1R_p$ (blue curve), and $r=2R_p$ (red curve) above 
the planetary surface, and normalized to the initial value for Case A (top; weak planetary field) and Case B (bottom; strong planetary field).}
\label{fig:f6}
\end{figure*}

In Figure~\ref{fig:f6}, we show the total mass flux through the three spheres as a function of time. The fluxes are normalized to their 
value at t = 00:00h, $F_{p0}=7\cdot 10^{-12}\;g\;cm^{-2}\;s^{-1}$. In Case A, the mass flux through the sphere at a hight of $h=2R_p$ 
reaches high negative values around t = 00:30h. This is a clear 
evidence for CME penetration to that height. However, the negative values of the mass flux for $h=1R_p$ are about 10 times smaller than the values for 
$h=2R_p$, while the mass flux measured at $h=0.5R_p$ remains positive throughout the simulation. The mass flux for $h=2R_p$ increases around t = 01:20h, 
due to the magnetosphere bouncing back from the compression by the CME, overshooting its original location and covering this sphere with higher density 
than the original value at that point. The initial magnetospheric compression is the cause 
for the mass flux increase near t = 01:00h for $h=0.5R_p$ and $h=1R_p$. In Case B, the mass fluxes of all three spheres initially increase due to the magnetospheric 
compression, but the flux through the $h=2R_p$ spheres changes its sign around t = 01:10h. Nevertheless, the absolute value of the mass flux is much lower than in 
Case A, indicating that while the CME can still penetrate to this depth, its impact is much weaker. In Case B, the sign of the total fluxes for 
$h=0.5R_p$ and  $h=1R_p$ remains positive, indicating that the CME does not penetrate to a height of $h=1R_p$ above the planetary surface 
($\sim$ 18 grid cells).

\subsection{Synthetic Observations of the Space Weather Event}
\label{sec:Observations}

When a CME reaches the Earth during a space weather event, it is observed and measured by satellites. 
Two of these satellites, WIND \citep{WIND95} and ACE \citep{ACE98}, are designed to continuously monitor the solar wind near Earth. 
A typical signature in these observations appears when a CME with a strong {\it negative} $\hat{z}$ component of the magnetic field reaches Earth. 
$\hat{z}$ is in the coordinate system associated with the Earth's magnetic dipole, which points in the {\it positive} $\hat{z}$ direction. 

\begin{figure*}[h!]
\centering
\includegraphics[width=4.25in]{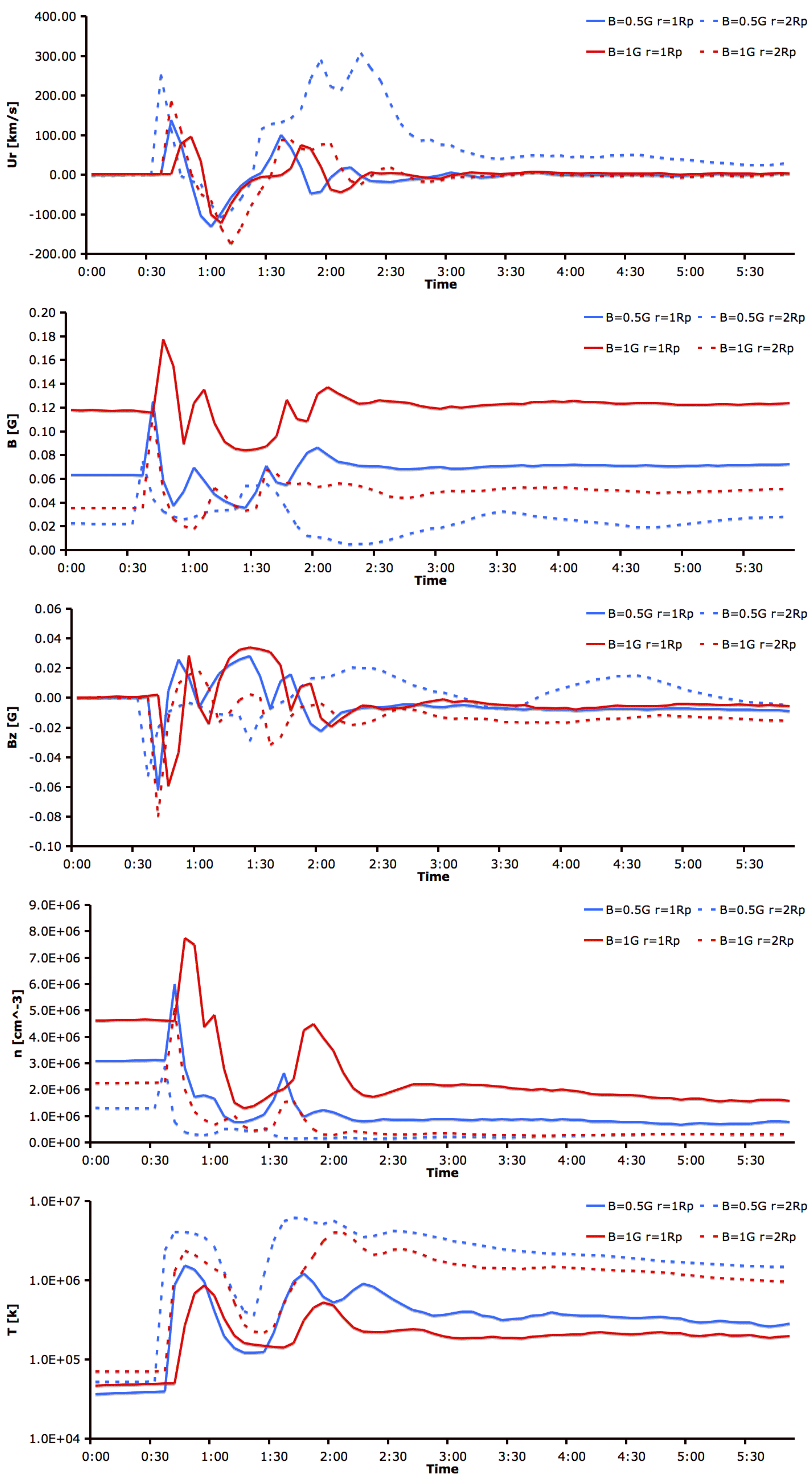}
\caption{Synthetic stellar wind values extracted from the simulation, mimicing measurements taken by imaginary satellites assumed to be 
located at the subsolar point at a height of 1$R_p$ (solid line) and 2$R_p$ (dashed line) above the planetary surface. Data are shown for Case A (blue) and Case B (red).}
\label{fig:f7}
\end{figure*}

In our simulation, we can extract the model plasma and magnetic field conditions at a particular time and location.  Figure~\ref{fig:f7} shows an extraction 
of the synthetic data from the simulation
along the trajectories of two imaginary satellites which are located at the planetary substellar point, one at a height of $h=1R_p$ and one at a height of $h=2R_p$. 
Until the CME shock arrival around 00:30h, the satellite located at $h=1R_p$ sees an ambient magnetospheric plasma, while the satellite located at $h=2R_p$ 
sees coronal plasma. The shock is clearly seen as a jump in all parameters and a decrease of the $B_z$ component to large negative values. 
The changes are not sharp as seen at real data taken near Earth, since here the differences between the ambient plasma and the CME plasma are not as marked 
as at 1~AU. The cold, less dense material seen in Figure~\ref{fig:f7} between 00:50h and 01:30h is not due to the CME cavity, 
but due to the planetary magnetosphere expanding and bouncing back after the initial impact by the CME. Between 01:30h 
and 02:20h, the magnetosphere shrinks again, causing another increase in all parameters.  After 02:20h, the synthetic data show a slow decrease in plasma 
number density and temperature as a result of a steady flow of the CME.  We do not recover the initial coronal parameters within the time domain of 
the simulation due to the fact that here we never turn the CME ``off'' as occurs naturally in the real event.  In 
this late, steady stage of evolution, the size of the planetary magnetosphere increases again due to the outflow from the mass source at the planetary 
surface, and the magnetospheic tail starts to grow as well. The dynamics described above can be seen in the online movies (see \S~\ref{sec:Interaction}).


\section{DISCUSSION}
\label{sec:Discussion}

In our study of a space weather event on a hot Jupiter, introduces unique properties 
of the interaction between the planetary magnetosphere of a close-in planet and a CME, based on a particular set of parameters. 
This interaction is clearly different than an interaction that takes place far from the star, as in the terrestrial case. 
we also discuss the amount of angular momentum the planet loses during the event.

\subsection{Properties of a Close-in CME-magnetosphere Interaction}
\label{sec:Unique}

\begin{figure*}[h!]
\centering
\includegraphics[width=5.in]{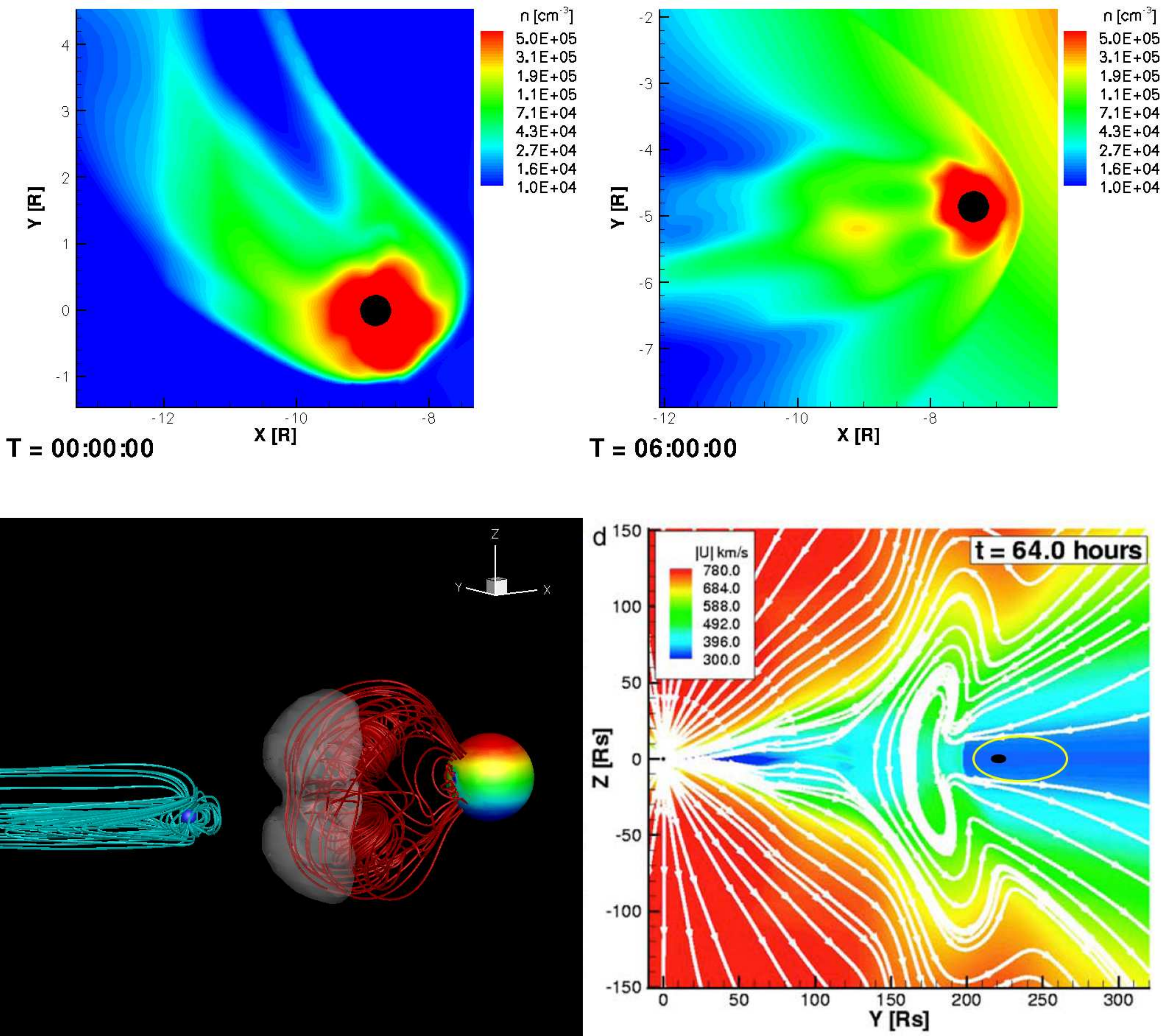}
\caption{Top: number density in the equatorial plane near the planet at t=00:00h (left) and t = 03:00h.
Bottom: comparison of the CME and the planetary magnetosphere sizes at different distances from the star --
left panel shows CME field lines (red), magnetospheric field lines (cyan), and an isosurface of $U_r=1500~km~s^{-1}$, which represents the CME front,
and right panel shows, for comparison, a meridional cut 
colored with contours of $U$ from a simulation of a solar CME at 1~AU (taken from \cite{manchester04}). The black ellipse represents a tentative size of the Earth's 
magnetosphere, while the yellow ellipse represents a tentative size of Jupiter's magnetosphere.}
\label{fig:f8}
\end{figure*}

We find the following features of the close-in planet-CME interaction. {\sl First}, 
the magnetospheric nose is initially located not at the subsolar point, but in the direction of the orbital motion 
\cite[also discussed in][]{Cohen11}. 
{\sl Second}, the fast planetary orbital motion results in a long comet-like magnetospheric tail \citep{Schneider98,Schneiter07}. 
Therefore, the magnetosphere is almost perpendicular in orientation to the direction of propagation of the CME, and the CME hits the 
magnetosphere from the side and not on its nose.  This orientation involves a contact surface between the 
magnetosphere and the CME that is much larger than in magnetosphere-CME interaction on Earth. {\sl Third}, about two 
hours after the CME onset time, the effect of the CME on the planetary magnetosphere becomes steady, when the magnetospheric 
orientation is changed from being perpendicular to the CME direction of propagation to being aligned with it. 
The magnetospheric nose at this stage is roughly at the substellar point, with a steady bow shock in front of it. 
This configuration resembles the steady, known magnetospheric configuration in our solar system. The initial, 
perpendicular, configuration and the steady, aligned configuration are shown in the upper panel of Figure~\ref{fig:f8}. 
{\sl Fourth}, by the (short) time the CME reaches the planet, it has not had time to be fully developed like 
CMEs in the solar system observed at Earth.  It does not expand much before it collides with the 
planetary magnetosphere.  Hence, it meets a body which is comparable in size to it (see bottom-left panel of 
Figure~\ref{fig:f8}).  In the solar system, the scale of CMEs is generally much larger than the planetary magnetospheres they meet 
by the time they propagate to large distances in interplanetary space (see bottom-right panel of Figure~\ref{fig:f8}), at least for the terrestrial planets. 
Even at Jupiter, the CME may not be much larger than the magnetosphere, but it can be considered as a plane ``wave'' that impacts the 
magnetospheric nose without being affected and modified itself by the interaction. In the case of close-in interactions studied here, the CME hits the 
magnetosphere from the side, it is slowed down and cools off as it expands around the magnetosphere. The CME is in effect split by the planet and ends up 
surrounding the planetary magnetosphere.  That the CME is affected to such an extent by the magnetosphere is a unique property of the 
close-in interaction.

\subsection{CME Impact on the Planet and Magnetospheric Configuration}
\label{sec:ImpactonPlanet}

Due to the unique interaction properties described above (\S~\ref{sec:Unique}), one might expect the impact of the CME on the planet to 
be stronger than in an interaction which takes place farther out in interplanetary space. Nevertheless, our simulation 
shows that even with a relatively weak planetary magnetic field, the CME does not penetrate more than $0.5R_p$ above the surface. 
This means that the planetary atmosphere is well-shielded from erosion by the dynamic pressure of the CME by even a quite modest planetary 
magnetic fields of 0.5 and 1.0~G. With a much weaker field, we would expect a more direct hydrodynamic interaction of the CME with thermal pressure 
of the planetary upper atmosphere \citep{Lammer07}. Such weak field is unlikely though, since the equatorial field strength of Jupiter is about 2.5G.

\begin{figure*}[h!]
\centering
\includegraphics[width=5.in]{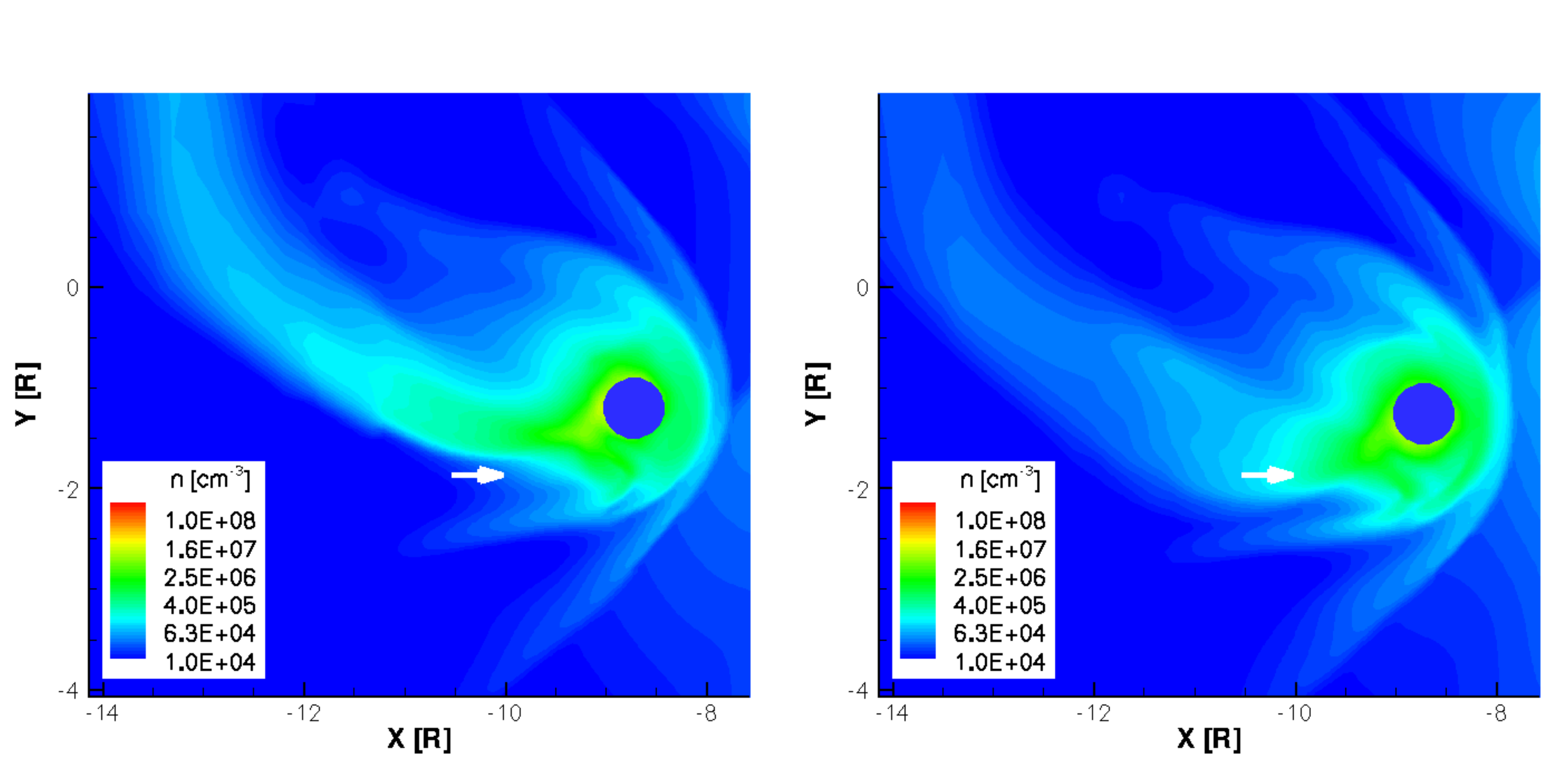}
\caption{Number density values, as in the top panels of Figure~\ref{fig:f10}, showing magnetotail plasma that gets `snapped' towards the 
planet between 01:20h (left) and 01:30h (right).  The arrows represent the direction of movement of the plasma.}
\label{fig:f9}
\end{figure*}

Here we studied a CME whose magnetic field orientation is exactly opposite to 
the planetary field. This topology leads to a continuous magnetic reconnection between the CME and the planetary fields at 
the the front of the magnetosphere, as well as behind the planet. In the case of the Earth, reconnected field lines at the dayside 
are dragged by the CME plasma to the night side, where they reconnect again.  During the time the field lines are dragged from the 
day to the night side, the magnetotail is stretched and energized, and then snapped due to the second reconnection.
This process drives a geomagnetic storm \citep{RussellKivelson95}. 
In our simulations, the CME completely surrounds the magnetosphere between 00:30h and 01:40h, 
leading to a detachment of a significant fraction of the magnetospheric tail.  During this disconnection, 
a blob of plasma is snapped towards the planet and is seen to travel radially inwards (as shown in Figure~\ref{fig:f9}). The average density of this plasma blob is 
$10^6~cm^{-3}$ and it is snapped through a distance of about $0.5R_\star$ within 10 minutes, so the average blob speed is 
$\bar{U}_{blob}\approx 450~km~s^{-1}$. From the simulation, we estimate the 
volume of this blob to be about $V\approx 10^{28}~cm^3$. These parameters 
give a total kinetic energy, $E_k=\frac{1}{2}\rho \bar{U}^2_{blob}\cdot V\approx 10^{25}~ergs$. This is three orders of magnitude higher than 
the typical energy of $10^{22}~ergs$ (or $10^{15}~J$) carried by the magnetotail fast flows towards the Earth during a geomagnetic substorm 
\citep{Pulkkinen02,Pulkkinen07,Tanskanen02}. However, this is about four-five orders of magnitude lower than the EUV energy flux 
a planet this size receives in 24 hours (the time-scae of the event). Nevertheless, the CME energy is deposited in an impulsive manner 
so that the effect on the planet could be significant due to the geomagnetic effects.

\begin{figure*}[h!]
\centering
\includegraphics[width=5.in]{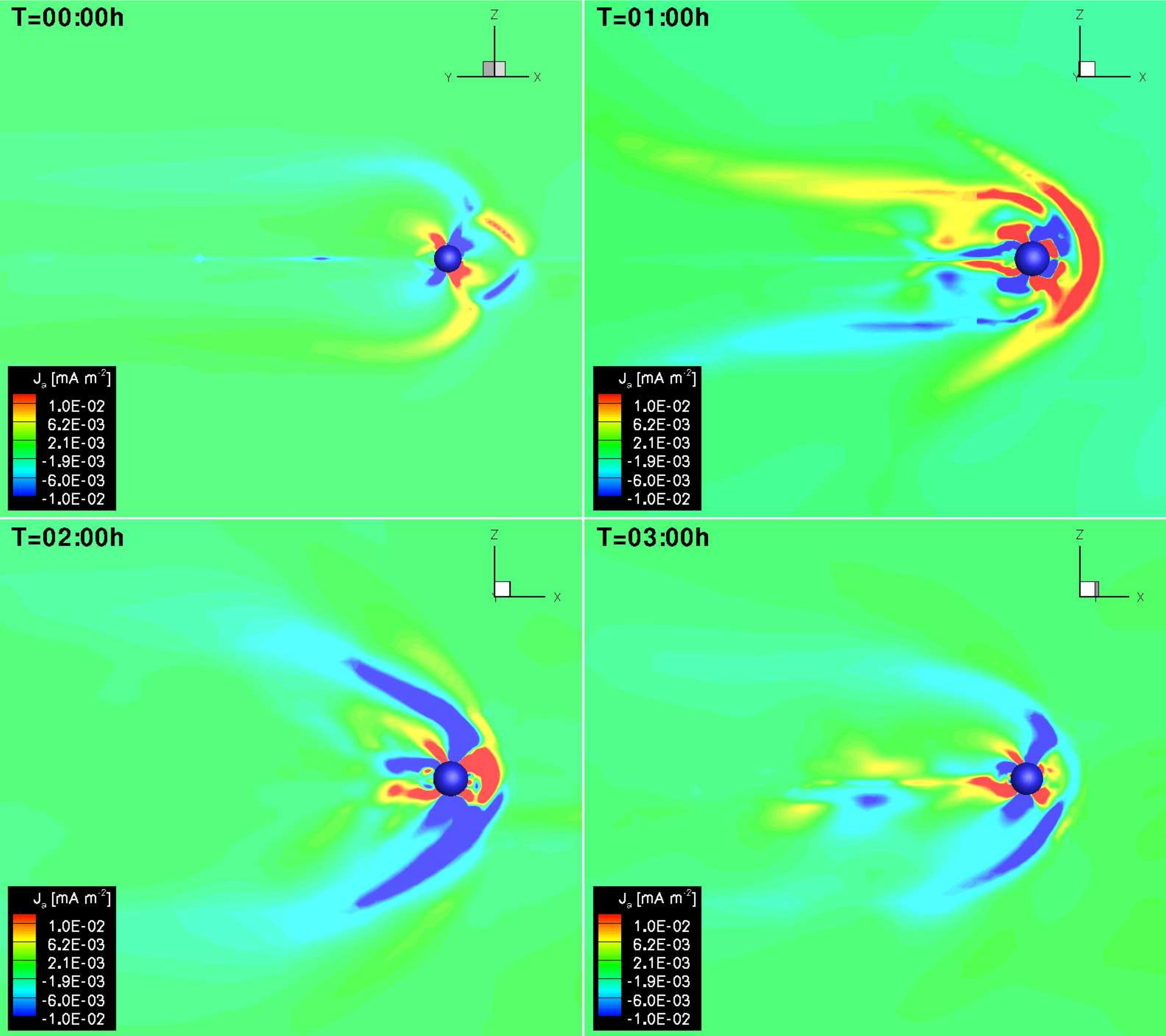}
\caption{Field-aligned currents are displayed over a meridional cut along the planetary magnetosphere for t = 00:00h, 01:00h, 
02:00h, and 03:00h. }
\label{fig:f10}
\end{figure*}

Figure~\ref{fig:f10} shows a meridional cut along the planetary magnetosphere at t=00:00h, 01:00h, 02:00h, and 03:00h, colored 
with the magnitude of the field-aligned current (current parallel to the magnetic field). The initial configuration resembles the Earth 
field-aligned current system, with the two tail lobes stretched behind the planet separated by a thin current sheet. As the CME 
hits the planet, the two lobes diverge and become more separated from each other in a topology that resembles ``Alfv\'en Wings'' \citep{Drell65,Neubauer80}. 
At t=03:00h, the trailing lobes have moved back toward each other. Alfv\'en wings are Alfv\'en waves generated around a body which moves in a magnetized plasma.  
These waves propagate away from the body, creating lobes with different properties from the surrounding plasma, and where the angle of these 
lobes depends on the Alfv\'enic Mach number.  For low Mach numbers, the lobes appear far apart while for high Mach numbers, the lobes are 
close to each other behind the body. 
The Mach number and the topology of the wings/lobes can have a significant effect on the magnetospheric response to the CME, the planetary 
polar cap potential, and the way energy is deposited from the CME into the magnetosphere \citep{Ridley07}.  Here there is a clear transition 
of the lobes during the event and the effect on the planetary magnetosphere due to this may be significant in the above context.  
However, a detailed numerical model for the magnetosphere itself is needed in order to study such a magnetospheric response to a CME and is beyond the scope of this work.

\subsection{Angular Momentum Loss During the CME Event}
\label{sec:AML}

The magnetospheric plasma is confined within the magnetosphere, and is moving with the planetary 
magnetic fields. Therefore, the disconnection of 
the magnetospheric tail and loss of associated plasma can involve a removal of angular momentum from the planet.   It is possible to compute the total angular momentum associated with the planet as the sum of the 
angular momentum of the planet and the angular momentum of the magnetospheric plasma. 
The angular momentum of the 
planet itself is:

\begin{equation}
J_p=\mathbf{r}\times\mathbf{p}=aM_pu_p,
\end{equation}
where $a=8.8R_\star=4.6\cdot10^{10}~cm$ is the semimajor axis, the planetary mass $M_p$, is assumed to be equal to Jupiter's, 
$M_J=10^{30}~g$, and the planetary angular velocity, $u_p\approx 100~km~s^{-1}$ (for orbital period of about 2 days at this distance). 
These parameters give an approximate planetary 
angular momentum of about $J_p=5\cdot10^{48}~g~cm^2~s^{-1}$. Based on the simulation initial result, we estimate 
the total angular momentum of the magnetospheric plasma, $J_M=\sum_i\rho_iV_i\mathbf{r}_i\times\mathbf{u}_i\approx10^{32}~g~cm^2~s^{-1}$, 
where the sum is over all the cells in the magnetosphere (excluding the planet itself). These cells are identified by a low temperature of $T<5\cdot10^5$~K. 
Later in the simulation, 
about half of the magnetospheric plasma is removed.  Taking into account that the plasma lost is less dense than that which remains attached to the planet, 
the planet loses about $J_M=5\cdot10^{31}~g~cm^2~s^{-1}$ during 
the event.  If there were ten such CME events per year, each of which is aimed directly at the planet, we 
obtain an angular momentum loss rate of $\dot{J}=10^{-16}~J_p~yr^{-1}$. This value is very small and cannot be significant source of planetary angular momentum 
loss for any plausible CME rate. 


\section{SUMMARY}
\label{sec:Conclusions}

We carry out an MHD simulation of a space weather event on a close-in planet in a manner similar to that for the impact of CMEs on Earth.
Our simulation yields complicated, non-trivial dynamics of a CME-magnetosphere interaction, introducing some unique properties that do 
not occur in the solar system. The main feature is that the initial orientation 
of the planetary magnetosphere, elongated to a comet-like tail by its motion in the stellar wind, is almost perpendicular to the CME direction 
of propagation. As a result, 
the CME is modified by the interaction. The resulting change in the CME topology and parameters 
complicates the dynamics so it is nearly impossible to be described by simple analytical relations. This complexity appears 
even in the case of the most simple initial magnetic topology we have used here. Despite its proximity to 
the host star, we find that the planet is well shielded from being eroded by the CME, even with a relatively weak intrinsic magnetic field of 0.5~G. 
We also find that the planetary angular momentum loss associated with a disconnection of part of the planetary tail is negligible compared 
to the total planetary angular momentum.  Our simulation suggests that the planetary magnetosphere can be significantly affected by the CME event, 
and that the energization of the planetary magnetospheric-ionospheric system might be much higher than in the Earth. It also suggests a transition 
in the magnetospheric Alfv\'en wings configuration during the event, as well as a rotation of the whole current system by 90$^\circ$. However, 
our simulation cannot provide such detailed information about the planetary properties; investigation of these aspects of the interaction 
requires a detailed numerical model for the planetary magnetosphere.


\acknowledgments
We thank an unknown referee for her/his many useful comments. We also thank KC Hansen, Alex Glocer, and Aaron Ridley for their useful 
comments on the magnetospheric response to the CME. 
OC is supported by SHINE through NSF ATM-0823592 grant, and by NASA-LWSTRT Grant NNG05GM44G.
JJD and VLK were funded by NASA contract NAS8-39073 to the {\it Chandra X-ray Center}.
Simulation results were obtained using the Space Weather Modeling
Framework, developed by the Center for Space Environment Modeling, at the University of Michigan with funding
support from NASA ESS, NASA ESTO-CT, NSF KDI, and DoD MURI.





\clearpage


\clearpage

\end{document}